\documentstyle[aaspp4,psfig]{article}
\def\ap{$\sim$}

\def\deg{$^\circ$}
\def\etal{{\it et al.}}
\def\g{$\gamma$-ray}
\def\co{{\it Compton Observatory}}
\def\min{$^{\prime}$}
\def\pt#1{$10^{#1}$}
\def\tpt#1{$\times10^{#1}$}
\begin{document}
To be published in Ap J, February 20, 1997 issue, Vol. 476

\title{An Intense Gamma-Ray Flare of  PKS\,1622$-$297
}
\author{J. R. Mattox\altaffilmark{1,2}
S. J. Wagner\altaffilmark{3}
M. Malkan\altaffilmark{4}
T. A. McGlynn\altaffilmark{2,5}
J. F. Schachter\altaffilmark{6}
J. E. Grove\altaffilmark{7}
W. N. Johnson\altaffilmark{7}
J. D. Kurfess\altaffilmark{7}
}
\altaffiltext{1}{Astronomy Department, Boston University, 725 Commonwealth Ave.,
Boston, MA 02215} 
\altaffiltext{2}{Former affiliation: Astronomy Department, University Of Maryland, and 
Universities Space Research Association}
\altaffiltext{3}{Landessternwarte Heidelberg--K\"onigstuhl, K\"onigstuhl,
D-69117 Heidelberg, Germany}
\altaffiltext{4}{University of California, Los Angeles, CA 90024}
\altaffiltext{5}{NASA/GSFC, Greenbelt, MD  20701}
\altaffiltext{6}{Harvard-Smithsonian Center for Astrophysics, Cambridge, MA 02138}
\altaffiltext{7}{Naval Research Laboratory, Washington DC, 20375}

\begin{abstract}
We report the  observation by the 
{\it Compton Gamma Ray Observatory} of a spectacular flare of 
radio source PKS\,1622$-$297.
A  peak flux of $(17\pm3)\times10^{-6}$cm$^{-2}$s$^{-1}$
(E $>$ 100 MeV) was observed.
The corresponding isotropic luminosity is
2.9\tpt{49}\ erg/s.
We find that PKS 1622$-$297 exhibits \g\ intra-day variability.
A flux increase by a factor of at least 3.6 was observed to occur in less than
7.1 hours (with 99\% confidence). Assuming an exponential rise, the
corresponding doubling time is less than 3.8 hours. A significant flux 
decrease by a factor
of \ap2  in  9.7 hours was also observed.
Without beaming, the rapid flux change and large isotropic luminosity
are inconsistent with the Elliot-Shapiro condition
(assuming that gas accretion is the immediate source of power for the \g s). 
This inconsistency suggests that the \g\ emission is beamed.
A minimum Doppler factor of 8.1 is implied by 
the observed lack of pair-production opacity (assuming x-rays are
emitted co-spatially with the \g s).
Simultaneous observation by EGRET and OSSE 
finds a spectrum adequately fit by a power law with photon index of $-$1.9.
Although the significance is not sufficient to establish this beyond doubt, 
the  high-energy \g\ spectrum appears
to evolve from hard to soft as a flare progresses.
\end{abstract}

quasars:general --- quasars:individual:PKS 1622$-$297 ---
gamma rays: observations

\newpage

\section{Introduction}
The Energetic Gamma Ray Experiment Telescope (EGRET) aboard the {\it Compton
Gamma Ray Observatory} is sensitive in the energy range 30 MeV to 30
GeV (Thompson \etal\ 1993). 
It has detected $\sim$50 AGN (Montigny \etal\ 1995, Thompson \etal\ 1995,
Mattox \etal\ 1996a) in the blazar class (by which we mean
 the ensemble of BL Lac objects, high polarization
quasars (HPQ), and optical violently variable (OVV) quasars). 
The absence of pair-production absorption in the EGRET spectra and
the fact that only sources which show parsec scale radio jet structure have been
identified as EGRET sources indicate
that the hard $\gamma$-rays are emitted
in a relativistic jet directed toward us. 

Most models feature inverse Compton scattering as the \g\ emission mechanism, but
there is not a consensus as to the origin of the low energy photons which are scattered. 
It has been suggested that 
they might originate within the jet as synchrotron
emission (Maraschi Ghisellini \& Celotti 1992;  Bloom \& Marscher 1993).
This is designated as the synchrotron self-Compton (SSC) 
process.
Another possibility is that the low energy photons come from outside of the jet.
This is designated as the external Compton scattering (ECS) process.
Dermer, Schlickeiser, \& Mastichiadis (1992) suggested that 
they come directly from an accretion disk around a 
blackhole at the base of the jet. It was subsequently  
proposed that the dominant source
of the low energy photons for scattering could be  re-processing of disk emission
by broad emission line clouds   
(Sikora, Begelman, \&  Rees 1994; Blandford \& Levinson 1995, 
Levinson \&  Blandford 1995; Levinson 1996).
Ghisellini \& Madau (1996) suggest that the dominant source of low energy photons for scattering
is broad-line-region re-processing of jet synchrotron  emission.
Hartman \etal\ (1996) find
that the  multiwavelength spectra of 3C 279 can be adequately fit
with either a 
SSC model or
an ECS model in both the high and low states.

The correlation of multiwavelength variability promises a means to
distinguish the SSC and the  ECS models. However, this is difficult
  because the sensitivity of
EGRET is insufficient to resolve variation on time scales shorter
than $\sim$1 week when blazars are
faint, and intense $\gamma$-ray\ flares are infrequent. 
Because of this, we proposed that a "quick look analysis" of EGRET data
be done to detect a \g\ flare in progress. This lead to our observation
of PKS\,1622$-$297.

PKS\,1622$-$297 has not received much attention previously
(being located in the Galactic center region, 
l=348.82\deg, b=13.32\deg).
It is not
cataloged by Hewitt \& Burbidge (1987, 1989). No optical polarization
measurement, nor search for rapid optical variability 
has been previously reported. However, the radio properties indicate
that it belongs to the blazar class.
A 5 GHz flux density of 1.92\,Jy and a
spectral index of $\alpha=+0.07$ ($S(\nu)\propto\nu^{\alpha}$)
were reported by  K\"uhr \etal\ (1981). 
Steppe \etal\ (1993) report  90\,GHz flux densities at three epochs
of 1.5, 1.8, and 2.0 Jy; and one 230\,GHz  observation at a flux density
of 1.0 Jy. Preston \etal\ (1985) report a VLBI
correlated flux density at 2.29\,GHz of 0.29\,Jy, 13\% of the total.
Impey \& Tapia (1990) report a
5 GHz radio polarization of 4.6\%.
It was optically identified by Torres \&  Wroblewski (1984) at 21 mag,
and by Saikia \etal\ (1987) at 20.5 mag.
A red shift of $z=0.815$\ is
reported in the PKS catalog (Wright \& Otrupcek 1990).
PKS\,1622$-$297 was detected by ROSAT during the sky survey
at a flux of 3.2$\pm$0.8 \tpt{-13} erg cm$^{-2}$s$^{-1}$ in the  
0.1-2.4 keV energy band  (Voges 1996).

\section{The Observation of PKS\,1622$-$297 with the {\it Compton Gamma Ray Observatory}}

PKS 1622$-$297 has been deeply exposed previously by EGRET.
A likelihood analysis (Mattox \etal\ 1996b)  
of the sum of EGRET exposure for the
first half  of the mission
(4/22/91 -- 10/04/94, a total exposure of 1.4\tpt9 cm$^2$ s) yields 
a 95\% confidence upper limit of
$0.10\times10^{-6}$cm$^{-2}$s$^{-1}$
(E $>$ 100 MeV). 
It was much brighter during our cycle 4 observation.
The exposure is given in Table 1 for each cycle 4 viewing period
(VP). 

\begin{table} 
\begin{center}
\begin{tabular}{|cccccc|} \hline
Viewing &\ \ \ \ \ \ \ \ Observation & Interval \ \ \ \ \ \ \ \   & Off-axis & EGRET Exposure & OSSE Exposure\\ 
Period  &  Gregorian Calendar & MJD  & Angle  & (cm$^2$ s)  &   (cm$^2$ s)  \\ \hline
421     & 6/06/95 -- 6/13/95& 49874.8 -- 49881.6 &  $14^\circ$ & 3.6\tpt7 &0 \\  
422     & 6/13/95 -- 6/20/95& 49881.6  -- 49888.7 & $14^\circ$ & 4.6\tpt7 &0 \\  
423     & 6/20/95 -- 6/30/95& 49888.8 --   49898.6  & $19^\circ$ & 3.7\tpt7 &0 \\  
423.5   & 6/30/95 -- 7/10/95& 49898.7 --  49908.5 & $2^\circ$ & 7.7\tpt7 &2.7\tpt8  \\  
424     & 7/10/95 -- 7/25/95& 49908.6 -- 49923.6&  $35^\circ$ & 0.7\tpt7 &0.9\tpt8 \\  \hline
\end{tabular}
\caption{
\baselineskip=11pt
GRO exposure for PKS 1622$-$297 during cycle 4. EGRET exposure is 
 for the energy selection E~$>$~100 MeV.
}
\end{center}
\end{table}

The position determined by likelihood analysis (Mattox \etal\ 1996b) with
the EGRET data (E$>$100 MeV,  VP 421.0 --- 423.5)
  is J2000 RA = 246.49\deg , DEC=$-$29.95\deg. 
The region of position
uncertainty is nearly circular with a radius of 15\min\ at 95\%
confidence.  
The significance of the detection is 25$\sigma$.
The $\gamma$-ray position estimate is  consistent with
PKS\,1622$-$297, differing by 6\min.
We use the method of Mattox \etal\ (1996a) to assess
the reliability of this identification. This method uses the number
density of potentially confusing sources which are as flat as
PKS\,1622$-$297 and as bright at 5 GHz, the fraction of $\sim$1~Jy
sources detected by EGRET, and considers where PKS\,1622$-$297 is
located in the EGRET position error ellipse. Because flat-spectrum
sources with a flux density of at least 1.9 Jy are rare (\ap1 per 500
square degrees), the identification is good. Assuming a prior probability
of 5.4\% that PKS\,1622$-$297 is a \g\ source (this is the 
fraction of blazars of this radio flux which EGRET detects, Mattox \etal\ 1996a),
the formal confidence of
a correct identification is 99.6\%. We show below that the
$\gamma$-ray source exhibits dramatic variability. The only type of
 identified  EGRET source which shows this type of
variability is the blazar type of AGN. Because
PKS\,1622$-$297 is the only bright
radio source with blazar properties near the EGRET position, 
the identification is even more secure than the formal confidence given above. 

\subsection{The EGRET Light Curve}

The observations shown in Table 1 have been analyzed to obtain an
EGRET light curve for the event energy selection  E $>$ 100 MeV. 
The exposure was binned according to the quality
of the EGRET sensitivity and the strength of the emission to obtain
as much time resolution as possible given the statistical limitations
imposed by the very sparse EGRET data. The EGRET sensitivity to PKS 1622$-$297
was limited because it was observed at a substantial off-axis angle
(except during VP 423.5)
and the response of EGRET at the time was about half of what it originally
was (Sreekumar \etal\ 1996 indicate a 
response in VP 423 of 42\%  of the original response
for E $>$ 100 MeV) due primarily to degradation of the spark-chamber gas.

Maps of counts and
exposure were constructed for adjoining time intervals.
A likelihood analysis (Mattox \etal\ 1996b)
of these maps was done to obtain flux estimates. The substantial flux of 
PKS 1622$-$253 (EGRET team publication, in preparation) was simultaneously fit
in order to include it in a background model. 
We have also looked carefully at nearby \g\ source 2EG J1631-2845 during our
observation. It is not significantly detected 
in the sum of VP 421.0 thru 423.5. The 95\% confidence flux upper limit  is 
$0.18\times10^{-6}$cm$^{-2}$s$^{-1}$ (E $>$ 100 MeV). 
A search for a short timescale flare of 2EG J1631-2845 found
only the expected indication of variability due to nearby 
PKS 1622$-$297.

\begin{figure}
\hbox{\psfig{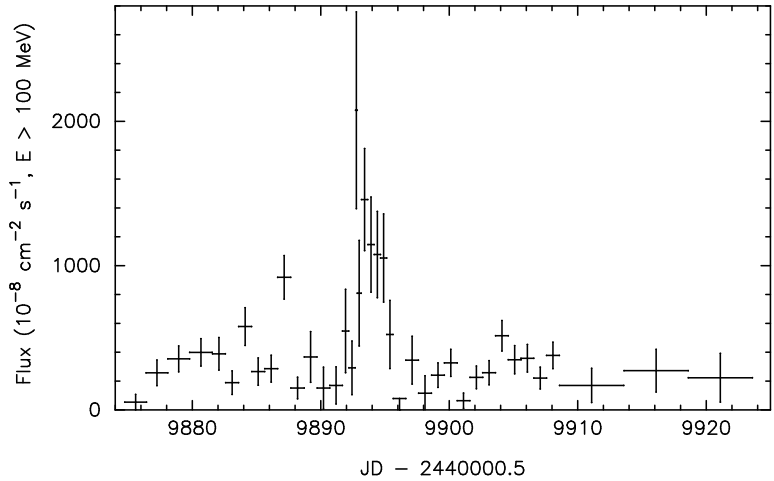}}
\caption{
\baselineskip=11pt 
The EGRET light curve for PKS 1622$-$297. The horizontal bars show
the extent of the time bins. The vertical bars indicate the ranges of the
68\% confidence flux estimates.
} 
\end{figure}

The result for PKS 1622$-$297 is shown in Figure 1.
The flux is observed to vary dramatically during this observation. A major
flare occurred at MJD 49894 (MJD $\equiv$ JD $-$ 2400000.5). The maximum
flux was $21\pm7 \times10^{-6}$cm$^{-2}$s$^{-1}$) for the 0.14 day bin
centered on MJD 49892.77. This small bin was 
chosen after the analysis of cumulative counts described below. It is
expected that this maximum flux is somewhat enhanced by selection. The
cumulative counts analysis described below
indicates that the apparent decrease in flux between this
bin and the subsequent bin has a  1.5\% probability of being statistical. 
A more conservative estimate of the maximum flux of
$(17\pm3)\times10^{-6}$cm$^{-2}$s$^{-1}$ (E $>$ 100 MeV)
results from an analysis of the combination of these two bins and the 
following bin (MJD 49892.7 -- 49893.7).

\subsubsection{An Analysis of Cumulative Counts}
The determination of the minimum 
time scale of variability of the high-energy $\gamma$-ray\ flux of
blazars is not straight forward 
because of the sparse EGRET statistics. The average
flux of PKS 1622$-$297 in cycle 4 divided by the average 
EGRET sensitivity is a count rate of
$\sim$1/hour. However, the flare at MJD 49894 to a flux $\sim$5 times this average potentially
provides the statistics 
 to detect variability on a short time scale.
Therefore, the data has been examined in detail.
The available information is summarized in the time history of cumulative counts
shown in Figure 2. 
This figure indicates that the onset of the MJD 49894 flare was  rapid.
There were 5 counts detected in the orbit at MJD 49892.72 and 7 counts
in the next orbit. In all eight preceding orbits, only 3 counts were
detected.
Unfortunately, the last of these eight 
orbits had only a small amount of exposure
due to limited availability of TDRSS satellites for telemetry downlink.
This increases the difficulty 
of determining the time and time scale of the flare onset
(intensifying the problem caused by sparse statistics).

\begin{figure}
\hbox{\psfig{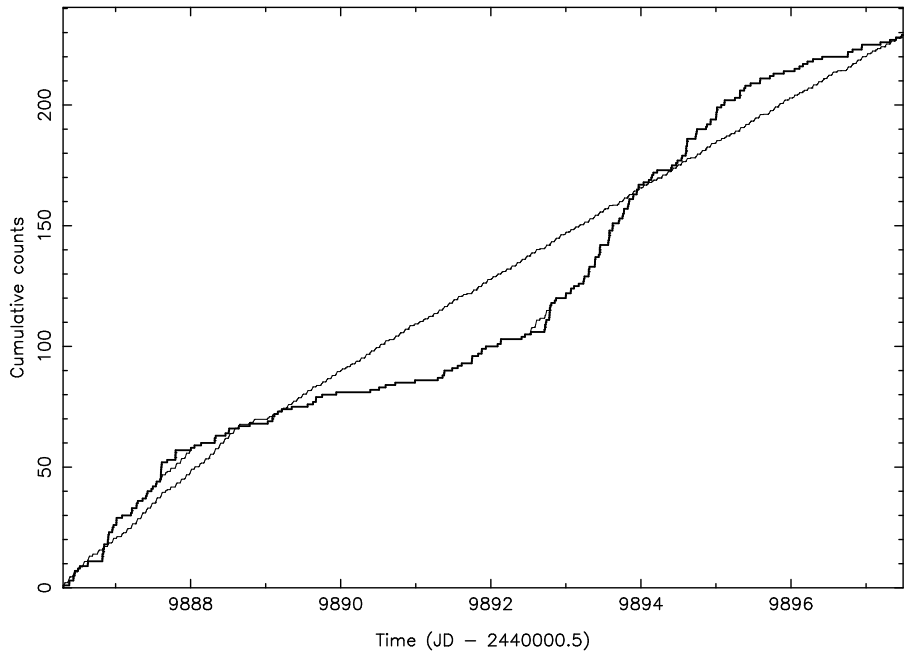}}
\caption{
\baselineskip=11pt 
The time history of cumulative EGRET counts 
for PKS 1622$-$297 is shown with the bold line. 
Events with E$>$100 MeV were selected from an
energy dependent cone which includes 68\% of the PSF.
The cumulative count time history expected for an invariant flux for the
entire interval is shown with
the continuous thin line. This is  an integral of the EGRET effective area
which was determined through an examination
of the  effective area of each trigger mode for each orbit.
The steps are due to Earth occultation during \ap1/2 of each 90 minute orbit.
Perturbations due to SAA passages and the occasional
unavailability of TDRSS satellites for telemetry downlink
are also apparent. The two short thin lines show the 
time histories  of invariant flux expected for shorter intervals
corresponding to KS tests described in the text.
} 
\end{figure}

We will begin by finding a time scale for which the
significance of a flux increase is beyond dispute.
The significance of variability is assessed by a
comparison of  the observed cumulative counts and those expected
for an invariant flux.  
The standard Kolmogorov-Smirnov (KS) test uses the maximum of the
absolute value of the
 difference between observed and expected cumulative counts.
Kuiper's variant of this test (Press \etal\ 1992) uses the sum of the maximum positive difference
and the absolute value of the maximum negative difference. For Figure 2,
the maximum positive and negative differences are 18.0 and 31.5
counts respectively. Confirming what is obvious by glancing at Figure 1,
the significance of variation using Kuiper's variant of the
KS test is \pt{-12} for the interval MJD 49889.5 --- 49897.5.

Now that variability has been established, we may confidently
seek variability in shorter intervals. 
We apply the standard 
KS test to the ten orbit interval described
above, MJD 49892.192 thru 49892.812.
The maximum deviation of the observed
counts from those expected for invariant flux is 8.5 counts
at MJD 49892.68 (where a cumulative 11.5 of 15 counts is expected
for  invariant flux). The corresponding
significance of variability is 6\tpt{-5}.  The high significance 
reported above for flux variation in the 8 day interval allows us to obtain
this result with not more than \ap10 trial intervals.

Now that we have established that a significant flux increase occurred at
MJD 49892.7, it is possible to examine the extent to which shorter
time scale  can be deduced. This can be done
without substantial trials to dilute the significance of the result
because we  examine a very limited number
 of sub-intervals within the interval for which
a significant flux increase has already been established.
We apply the KS test to the five orbits ending with the two bright orbits (MJD 
49892.517 -- 49892.812). The expected time history  of the invariant flux 
for this interval is shown with a short, thin line in Figure 2.
The maximum deviation of the observed
counts  is 5.7 counts
at MJD 49892.68 (where a cumulative 6.7 of 13 counts is expected
for  invariant flux, and 1 is observed). The 
standard KS test indicates a flux increase with a 
significance of 9\tpt{-3}. This implies a flux increase in less than
7.1 hours with 99\% confidence. 

It is  conventional to report the observed time for a  doubling 
of the flux. However, it is not possible to measure a  doubling time
directly in this instance because of limited statistics.
A likelihood analysis for the 3 day interval before the flare (9889.7 -- 9892.7)
yields a flux of  $2.7\pm0.9\ \times10^{-6}$cm$^{-2}$s$^{-1}$ (E $>$ 100 MeV).
This flux uncertainty is the 1$\sigma$ error.
The conservative peak flux obtained above is $17\pm3$.
An integration of the probability distribution function for the factor by 
which the flux increases obtained from
the ratio of these two Gaussian distributions (Papoulis 1991),
$$p(r)=\int_0^\infty G_h(r y)G_l(y) y dy,$$
indicates with 95\% confidence that the flux increase is greater than a
factor of 3.6.
Assuming an 
exponential rise, the corresponding upper limit on the doubling time is 3.8
hours.



A rapid flux decrease is also noted. A  KS test shows variability (with 
99.996\% confidence) for the
interval MJD 49886.8 --- 49888.6. Two brief intervals of high flux
are apparent at MJD 49886.85 and 49887.61.
The second is apparent
with a KS test at 99.91\% confidence for the interval MJD 
49887.593 --- 49887.996.  
The expected time history  of the invariant flux 
for the latter interval is also shown with a short, thin line in Figure 2.
This interval indicates a flux decrease in less
than 9.7 hours from a flux of $9\pm3$ (during MJD 49887.468 --  49887.624) 
to $3\pm2\ \times10^{-6}$cm$^{-2}$s$^{-1}$, E $>$ 100 MeV (during MJD 49887.660  --  49887.946). 

It is possible that the time scale of the flux change is not resolved
with EGRET, and is much faster\footnote{
Mattox (in preparation) finds that five of the seven events in the 
orbit at the end of the brightest part the MJD 49894 flare
occurred in the first 10 minutes of that orbit
which corresponds to only 14\% of the exposure of the orbit.
A KS test indicates a flux
decrease within this exposure interval 
(48 minutes long) with a confidence of 99\%.  However because of a 
a substantial number of potential trials, this is not a definitive result.}. 
If so, the observations of the proposed GLAST satellite
(Michelson \etal\ 1996) will be of great interest with
\ap10 times the effective area of EGRET.
The possibility of
a \g\ flux change in less than \ap1 hour is very interesting. 
This would severely constrain the size (or the Doppler
factor) of the \g\ emission region. 

\section{The Spectrum of PKS 1622$-$297 }

The ToO  pointing in VP 423.5 lead to an
OSSE detection (Kurfess \etal\ 1995). 
OSSE observes in the 50
keV to 10 MeV energy range.  Its small field of view limits its observations
to one object at a time so that coordinated observations of specific EGRET
targets must be planned in advance or arranged via a ToO.
Subsequent to the ToO pointing, OSSE was able to continue monitoring PKS 1622$-$297 during
VP 424 with the reduced sensitivity provided by
two of the four detectors at an off-axis angle of  2.3\deg. 

\begin{figure}
\hbox{\psfig{figure=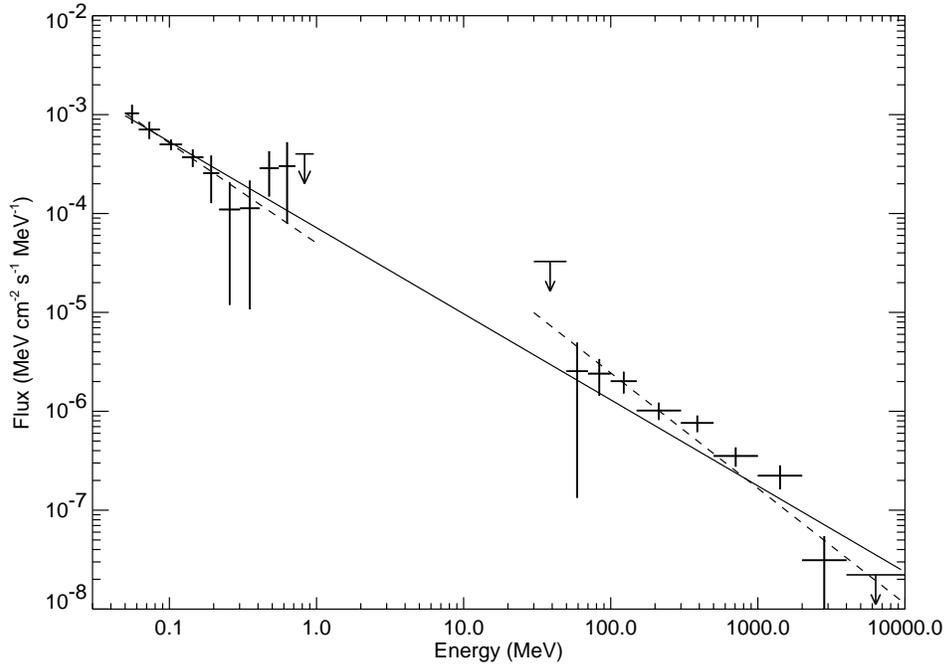,width=5.in,angle=90}}
\caption{
\baselineskip=11pt 
The spectrum of PKS 1622$-$297 for VP 423.5. The observed fluxes are shown with the crosses. The horizontal bars show the extent of the energy bins. 
The vertical bars indicate the 1$\sigma$ ranges of the flux estimates. 
The first 10 energy bins are from OSSE, the last 10 are from EGRET. 
The best power law fit to all the data is indicated by the solid line.
Power law fits to the data of each instrument alone are  
indicated by the dashed lines.} 
\end{figure}

We have obtained an OSSE/EGRET spectrum
for VP 423.5 which is strictly simultaneous.
The counts matrix and the response matrix for the
standard EGRET spectral analysis  (Nolan \etal\ 1993) were converted
to the XSPEC format\footnote
{The conversion program is available: send e-mail to the author 
(Mattox@bu-ast.bu.edu), or contact personnel at the \co\ Science Support 
Center (WWW URL http://cossc.gsfc.nasa.gov). XSPEC is described at
http://legacy.gsfc.nasa.gov/docs/xanadu/xspec/u\_manual.html.}. 
The XSPEC program was
used to simultaneously analyze the EGRET and OSSE results. 
The result is shown in Figure 3.
The OSSE data alone are well represented (reduced $\chi^2=0.74$)
by a power law with a photon spectral
index of $-2.0\pm0.2$.
The best power-law fit to EGRET data alone has a photon spectral
index of $-2.2\pm0.1$.
With a reduced $\chi^2=1.64$, the EGRET fit is adequate but not compelling.
It appears to break gradually to a steeper spectrum
with increasing energy. Similar convexity is  apparent in the EGRET 
spectra of several other blazars (see Montigny \etal\ 1995), especially
 1633+382 (Mattox \etal\ 1993). 

Although these OSSE and EGRET spectra have the same index, the normalization
of the EGRET spectrum is a factor of 5 larger than that for the 
OSSE spectrum at 10 MeV. When the OSSE and EGRET data are fit together with
a single power law, the result shown with the solid line
in Fig. 3 is obtained.
The photon spectral
index is $-1.87\pm0.02$.
The fit is not excellent, but acceptable with a reduced $\chi^2=1.14$. 
For comparison, the reduced $\chi^2$ for the total OSSE and EGRET data for
the two separate power-law fits described above is 0.86. 
The F test indicates that the  separate  fits do not offer a significant 
improvement (with a 14\% chance probability).
It is clear that the break at a few MeV to a harder x-ray spectrum 
which has been reported for 4 of the 5 blazars previously detected by 
both EGRET and OSSE (McNaron-Brown \etal\ 1996, observations were simultaneous
for one object, only contemporaneous  for the others)
is not apparent for PKS 1622$-$297.

\subsection{Spectral Evolution}

We  examined the EGRET data for evidence of spectral evolution
during the flare. 
The result is shown in the 
scatter plot of Fig. 4. 
We have done a likelihood analysis of the flux 
for two \g\ energy intervals: the  100 -- 300 MeV and E $>$ 300 MeV. 
We analyzed time intervals that were as short as possible to
provide sensitivity to spectral evolution
on the short timescales seen for flux changes. However, each interval had
to be long enough to provide sufficient statistics. Thus, we used
shorter time intervals when the flux was large.
The intervals were primarily
formed by combining the time bins shown in Figure 1.
However, the b and c intervals correspond to a single interval in Figure 1.
This interval was split after
an examination of a scatter plot of event energy verses
time indicated that the emission appeared much harder during the
first part of the interval. Interval b corresponds to 
the  interval of high flux apparent in Figure 2 at MJD 49886.85.
The exact time intervals can be obtained
from the caption of Figure 4.

\begin{figure}
\hbox{\psfig{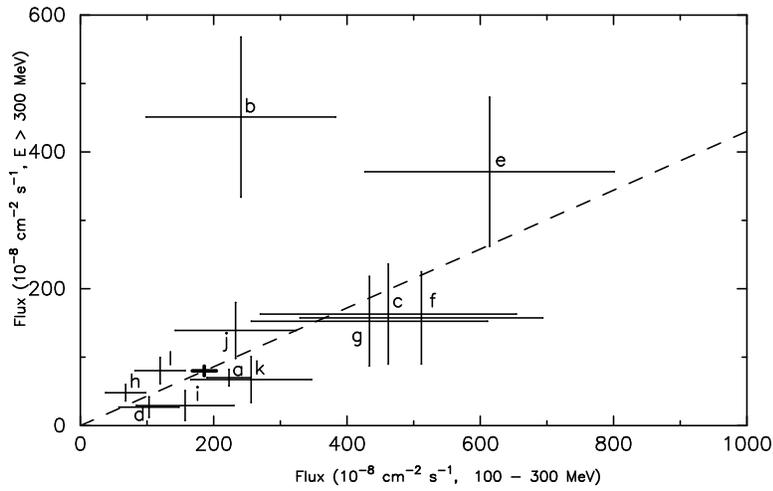}}
\caption{
\baselineskip=11pt 
A search for the 
spectral evolution of PKS 1622$-$297. Each time interval is
represented by a cross. The  bars indicate the 68\% confidence ranges of the
 flux estimates.
The intervals are labeled sequentially with a -- l.
The interval boundaries
are MJD 49876.42, 49886.65, 49887.28, 49887.66, 49892.674, 49893.660, 49894.646, 49895.632,
49902.60, 49903.60, 49904.60, 49905.59, and 49908.57. The bold cross 
indicates the  fluxes found for an analysis of
the sum of the entire data set.
The dashed line through this  bold cross indicates where  fluxes would be expected
to lie if the spectrum were invariant.} 
\end{figure}

Assuming an invariant spectrum, the ratio of these
fluxes is expected to be consistent with that found for an analysis of
the total exposure.
Under this assumption, 
it is appropriate to assume the  flux ratio observed in the total exposure and
fit a single intensity parameter for each time interval.
A $\chi^2$ statistic is then  obtained by summing the deviation
of both energy ranges for each time interval. 
The result for all 12 time intervals
($\chi^2=15.2$ for 12 degrees of freedom)
indicates  spectral variation with only 77\% confidence. We have also
done the analysis with 3 time bins (49876.2 -- 49892.7,
49892.7 -- 49895.6, 49895.6 -- 49908.6) and see colors  consistent
with an invariant spectrum.

It is interesting that all three intervals
which are at a peak in Fig.1 (b, e, \& j) are harder than the average. 
This is consistent with the  report of Mukherjee \etal\
(1996) of a marginally significant indication that the EGRET
spectrum of PKS 0528+134 was harder during an interval of high emission than
it was during lower emission level intervals.
The hardest interval is b for which $\chi^2=6.9$ (for 1 degree of freedom). 
This corresponds to a 2.4$\sigma$ deviation from an invariant spectrum.
The most energetic  PKS\,1622$-$297 event
(9.7$\pm$1.7 GeV) occurred during this interval at MJD 49886.897. 
We note that the spectrum appears to evolve from hard to soft (clockwise
in Figure 4) for all 3 peaks. Further observation is required to confirm
this indication  of spectral evolution of the \g\ emission of blazars.

\section{Multiwavelength Observations of PKS 1622$-$297 }
The large flux of
PKS 1622$-$297 was apparent to the EGRET team during a routine "quicklook"
analysis of the VP 421 data. 
The news of the detection was conveyed to us on  6/14/95
and triggered our ToO program. 
Fortuitously, this followed the first week of a 5 week observation of the Galactic center region which included
PKS 1622$-$297, and this source was opposite the sun. 
The detection was announced in an IAU Circular
(Mattox \etal\ 1995). 
We also notified directly a group of multiwavelength observers who had been
previously enlisted. 
Unfortunately, the nearly anti-solar position of PKS\,1622$-$297
at the time prevented x-ray observation by either ROSAT or ASCA, and
the {\it Rossi} XTE satellite had not yet been launched. 
Detailed multiwavelength results
including multiwavelength spectra will be published latter
(Mattox in preparation).
We will also report subsequently on possible BATSE and COMPTEL detections
and a possible correlation between the EGRET and OSSE flux in VP 423.5.

Extensively  optical monitoring of PKS 1622$-$297
commenced when we learned of the  EGRET detection.
We found  that PKS 1622$-$297 was 3 magnitudes brighter 
than its quiescent state, and that the brightness was observed to vary by 
as much as 150\% in less than 24 hours.
Thus, PKS\,1622$-$297 displays optical IDV (intra-day variability)
as do all other blazars
detected by EGRET which have been observed frequently enough for
IDV to be detected (Wagner 1996).
Several radio observations  in the mm band occurred  which showed a  flux density
significantly higher than previously observed.
It was also detected by IUE (Bonnell \etal\ 1995) during observations
on July 1, 5, 6, 8, and found to be variable.
In collaboration with Alan Marscher, a VLBA observation of PKS 1622$-$297 
was made on 7/25/95. Subsequent observations will be used to search for a new radio component corresponding to the \g\ flare.

\section{Discussion}

The maximum EGRET flux observed from PKS\,1622$-$297
$(17\pm3 \times10^{-6}$cm$^{-2}$s$^{-1}$, E $>$ 100 MeV)
was a factor of 2 larger than that of the Vela Pulsar which is normally
the brightest EGRET source. It is
a factor of 4 larger than the flux of 3C 279 at the peak of 
the 1991 flare (Kniffen \etal\ 1993).
A peak energy flux of 1.6\tpt{-8} ergs cm$^{-2}$s$^{-1}$
for the observed energy
range 30 MeV to 10 GeV is obtained for PKS\,1622$-$297 assuming a power law
spectrum with photon index of $-$1.9.
The corresponding isotropic luminosity is 2.9\tpt{49}\ erg/s 
assuming
a Friedmann universe with $q_o=1/2$
and  $H_o = 75\ {\rm km\ s^{-1} Mpc^{-1}}$. 
For this luminosity
to be less than the standard Eddington limit, the central black hole
mass must exceed 2\tpt{11} M$_\odot$.
However the standard Eddington limit does not pertain to the 
\g\ luminosity because the Compton scattering cross section in the Klein-Nishina regime
is much
smaller than the Thomson cross section (Dermer \& Gehrels 1995, 
Pohl \etal\ 1995). 
Using the expression of Dermer \& Gehrels (1995)   for the 
cross section in the Klein-Nishina regime for an EGRET flux of
$70\times10^{-6}$cm$^{-2}$s$^{-1}$ (E $>$ 30 MeV),
the lower limit for the mass of the PKS\,1622$-$297
black hole is 8\tpt{8} M$_\odot$. The corresponding limit on the
Schwarzschild radius is $R_g > 2.5$\tpt{14} cm. 

We apply the 
"Elliot-Shapiro argument" (Elliot \& Shapiro 1974; Dermer \& Gehrels 1995;
 Pohl \etal;
Dondi \& Ghisellini 1995, we note that they err in assuming the 
Thomson cross section for Compton scattering
in the Klein-Nishina regime)
to establish that the \g\ emission must occur in a relativistic jet
by showing that a contradiction follows otherwise.
Under the assumption that the \g\ emission is unbeamed, the observed
upper limit on the doubling time of $\tau = 3.8$ hours implies that the
extent of the emission region is less than
$c\tau/(1+z) = 2.2$\tpt{14}\ cm. The fact that this upper limit
is less than lower limit for $R_g$ suggests that
the \g\ emission  occurs in a medium which is moving relativistically
with a lateral extent
less than $c\tau\delta/(1+z) = 2.2$\tpt{15}($\delta/10$)\ cm,
and an extent along the jet of
less than $c\tau\delta^3/(1+z) = 2.2$\tpt{17}$(\delta/10)^3$\ cm,
where 
$\delta\equiv\gamma^{-1}(1-\beta cos \theta)^{-1}$
is the relativistic Doppler factor, $\theta$ is the angle between the jet axis and the
line of sight, and $\gamma\equiv(1-\beta^2)^{-1/2}$, where 
$\beta$ is the bulk velocity in units of the speed of light.
This argument assumes that the accretion of optically thin material 
is the immediate source of power
for the  \g s. This may not be the case. Accretion energy could be
stored in the rotation of a black hole, or in 
magnetic fields to power the \g\ flare; or the \g\ flare could result
from the accretion of a star.
Also, for a Kerr blackhole, emission could occur as close as $R_g/2$.

Another argument for \g\ emission in a relativistic jet can be made from the
lack of $\gamma-\gamma$ absorption in the \g\ spectrum. If the emission does not 
take place in a relativistic jet,
the \g\ emission would be absorbed by \g/x-ray pair production.
Mattox \etal\ (1993) derived the expected optical depth under the
assumption that the x-rays are produced co-spatially with the \g s. Their
expression was in error due a misunderstanding of the definition of 
luminosity distance. The corrected expression 
(assuming a Friedmann universe with $q_o=1/2$) is
$$\tau=2\times10^3 (1+z)^{(2\alpha)} (1+z-\sqrt{1+z})^2
 h_{75}^{-2} T_5^{-1} {F_{keV}\over\mu Jy} \Bigr({E_\gamma\over GeV}\Bigl)^{\alpha}
\ \ \ \ \ \ \ \ \ \ \ \ \ \ \ \ \ \ (1)$$
where $T_5$ is the time scale of variation in units of $10^5$ seconds, 
 $F_{keV}$ is the observed x-ray flux at a keV,
$\alpha$ is the x-ray energy spectral index, $F(\nu)\propto\nu^{-\alpha}$,
 ${E_\gamma}$ is the \g\ energy, and
 $h_{75}=H_o/(75\ {\rm km\ s^{-1} Mpc^{-1}})$. 
The luminosity distance 
error of Mattox \etal\ (1993) also affected their expression for the lower
limit of the Doppler factor. The corrected expression is
$$\delta\ge \Bigl[ 5\times10^{-4} (1+z)^{-2\alpha} (1+z-\sqrt{1+z})^{-2}
 h_{75}^{2} T_5 \Bigl({F_{keV}\over\mu Jy}\Bigr)^{-1}
\Bigr({E_\gamma\over GeV}\Bigl)^{-\alpha} \Bigr]^{
-(4+2\alpha)^{-1}}
\ \ \ \ \ (2)$$
Assuming  a spectral index ($\alpha=0.7$) typical of blazars,
the ROSAT sky survey  flux
for PKS\,1622$-$297 is $F_{keV}= 0.054 \mu Jy$.
The corresponding optical depth from equation (1) is 
$$\tau=330 \Bigr({E_\gamma\over GeV}\Bigl)^{0.7}$$
No indication of such absorption is apparent in the spectrum of Figure 3.
The lower limit for the Doppler factor from equation (2) (for
$h_{75}=1$, and $\tau<1$ for $E_\gamma=3$ GeV) is $\delta\ge 8.1$.
We note that the  x-ray observation was not simultaneous. If it were a factor
of  3 lower during the \g\ flare, the lower limit on $\delta$ would
decrease to 6.6.

\section{Conclusions}
We report the brightest, most luminous \g\ blazar ever detected. 
It shows the most rapid \g\ flux change yet seen for any blazar, with
a flux doubling time of less than 3.6 hours. 
This is the first observation of
\g\ IDV for a blazar.
We show that the Elliot-Shapiro argument and
the lack of $\gamma-\gamma$ absorption of \g s
indicate that
the  emission occurred in a relativistic jet.

Our results
  are not inconsistent with the  prediction of Ghisellini \& Madau (1996)
of an invariant \g\ spectrum throughout the flare. 
If their model of Compton scattering of synchrotron photons 
reprocessed in the broad-line region
pertains, the lightcurve shown in Fig. 1 might indicate the radial
profile near the jet of the density of the broad-line region.
Romanova and Lovelace (1996) have also developed a model for \g\
blazar flares which is consistent with our data. 
In their model, flares occur
when electrons  are accelerated by a shock in the jet.

Our results do not demonstrate the 
soft to hard spectral evolution expected 
for a \g\ emitting jet component emerging from a \g\ photosphere
(Blandford \& Levinson 1995, Levinson \&  Blandford 1995; Levinson 1996).
However, a comparison of a detailed calculation based on their model  
to these data must be done before quantitative limits can be placed on
their model.
Although the statistics of this exposure are not sufficient 
to establish this beyond doubt, 
the  high-energy \g\ spectrum appears
to evolve from hard to soft as a flare progresses, 
the opposite of the prediction of this model.

\vskip 1cm \centerline{Acknowledgments}

We are grateful to the EGRET team for the timely processing of telemetry
and data to obtain the "quicklook" detection of PKS 1622$-$297, and to
anonymous referees for useful comments.
J. Mattox  acknowledges support from NASA Grants NAG5-2833 and NAGW-4761.

\end{document}